\documentclass[journal=langd5,manuscript=article,layout=twocolumn]{achemso}

\usepackage[version=3]{mhchem} 

\newcommand{\dC}{$^{\circ}$C}

\newcommand{\EPSB}[5]{																	
 		\begin{figure}[H]																	
 		\centering
 		\includegraphics[width=#2]{#1}
 		\caption[#5]{\small \label{#3} #4}
 		\end{figure}
 		}

\newcommand{\EQ}[3]{		
		\begin{equation}
		\label{#1}
		#2\;\;#3
		\end{equation}
		}
\newcommand{\eg}{e.\,g.~}	
\newcommand{\ie}{i.\,e.~}	

\newcommand{\etalol}{\emph{et al.}} 

\newcommand{\cytc}{cytochrome \emph{c}~}
\newcommand{\cytcol}{cytochrome \emph{c}}	
\newcommand{\Ang}{\tn{\,\AA}} 

\newcommand{\tn}[1]{\textrm{#1}} 

\usepackage{color}
\usepackage{hyperref}

\author{Sebastian T. Moerz}
\author{Patrick Huber}
\email{sebastian.moerz@inm-gmbh.de, patrick.huber@tuhh.de}
\affiliation[Saarland University]
{Experimental Physics, Saarland University, D-66041 Saarbr\"ucken, Germany}
\alsoaffiliation[Hamburg University of Technology]
{Institute of Materials Physics and Technology, Hamburg University of Technology (TUHH), D-21073 Hamburg-Harburg, Germany.\\
Pre-print version of the article.\\ For the published article see \color{blue}{\href{http://pubs.acs.org/doi/abs/10.1021/la404947j}{Langmuir 30, 2729 (2014)}}, \href{http://dx.doi.org/10.1021/la404947j}{doi:10.1021/la404947j}\color{black}. }

\title[Protein Adsorption into Mesopores]
  {Protein Adsorption into Mesopores: A Combination of Electrostatic Interaction, Counterion Release and Van der Waals Forces}

\begin{document}

\begin{tocentry}

\includegraphics{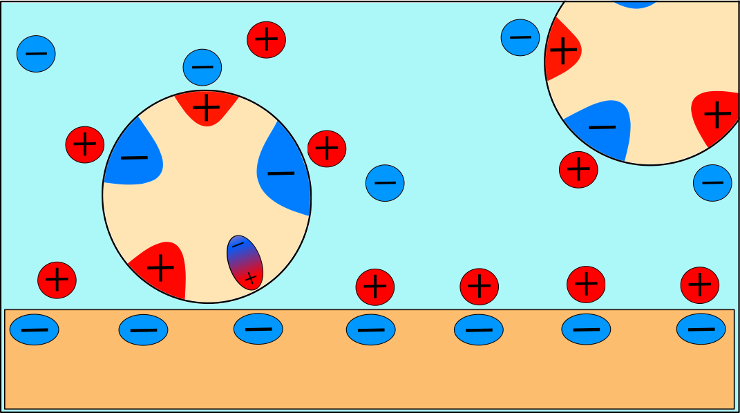}

\end{tocentry}


\begin{abstract}
 
Bovine heart cytochrome \emph{c} has been immobilized into the mesoporous silica host {material} SBA-15 in both its native folded and urea-unfolded state. The comparison of the two folding states' behavior casts doubt on the commonly used explanation of \cytc adsorption, \ie the electrostatic interaction model. A detailed investigation of the protein binding as a function of pH and ionic strength of the buffer solution reveals the complex nature of the protein-silica interaction. Electrostatic interaction, van der Waals forces and entropic contributions by counterion release each contribute to adsorption on the silica pore walls.

\end{abstract}

\section{Introduction}

Over the course of the last two decades, protein adsorption in mesoporous materials has been intensively studied and plenty of technical applications have been envisaged. Some proteins show an enhanced stability against denaturating conditions -- chemical as well as thermal -- and retain or even increase their electrochemical activity when immobilized in silica mesopores\cite{washmon-kriel_1999}. More trivially, since microorganisms like bacteria or fungi are far too large to penetrate mesoporous structures, encapsulated proteins are well protected from biological decomposition. This opens a wide field of biochemical applications that employ the enzymatic activity of proteins under conditions which would otherwise destroy the enzymes.
Other interesting applications arise from the fact that not all polypeptides adsorb equally well on all surfaces \cite{Quinn2008, Evers2011, haehl2012, Langdon2012}. Thus, fractionation of complex protein solutions and applications in chromatography should be feasible with customized mesoporous host materials.
A third and rather promising field of interest is the use of porous materials as novel devices for controlled \emph{in-vivo} drug release. Several researchers suggest utilizing this by loading a porous structure with drugs or enzymes and then injecting the loaded particles into living cells where the physiological conditions cause a release of the drugs into the cytosol. The feasibility of this application is demonstrated in literature\cite{Ho_2008} and especially by Slowing \etalol. \cite{slowing_2007}. The latter group loaded MCM-41 silica particles with a fluorescent protein which was subsequently released into the cytoplasm of human cervical cancer cells (HeLa-cells).

As a general rule of thumb, protein adsorption appears to be non-specific and can be reversible under specific conditions. Interestingly enough, despite the vast amount of research in this field, there is to date no consensus in literature what type of interaction dominates the adsorption of biomolecules on inorganic surfaces. It is the main motivation of this study to compare two of the most common models of protein adsorption and to highlight their advantages and shortcomings.

A widely used model system is the adsorption of \cytc into the mesoporous silica material SBA-15. There exists a vast amount of studies concerning the adsorption of \cytc on SBA-15, including works by Deere\cite{Deere_2002}, Hudson\cite{Hudson_2005}, Miyahara\cite{miyahara_2006_2}, Vinu\cite{Vinu_2004} and Zhang\cite{zhang_2007}. Reviews were published by Hartmann\cite{Hartmann_2005} and Zhao\cite{zhao_2011}. 
The behavior of this system is commonly attributed to simple electrostatic attraction. Nevertheless, all these studies to date focus on folded, \ie native \cytc exclusively. Here, we propose a comprehensive analysis of both folded and unfolded \cytc to gain new insights into the interactions which drive protein adsorption.

\subsection{Electrostatic Interaction Model}
\label{electrostatic_lit}

The highest pore loadings, \ie the amount of bound protein per gram of the silica material, are often found close to the isoelectric point where the overall charge of the protein is zero. For example, Vinu \etalol \cite{Vinu_2004} studied the pH-dependent adsorption of horse heart \cytc on SBA-15 in 25\,mM buffer solutions. They examined the pH-dependency between pH\,3 and pH\,$10.6$. The highest pore loading was observed at pH $9.6$ which is only slightly below the pI of \cytcol.

This is often interpreted in terms of a balancing between an attractive protein-wall interaction and protein-protein repulsion\cite{miyahara_2006_2}.\textsuperscript{,}\cite{Hartmann_2005} The loss of electrostatic repulsion between the molecules at their pI facilitates the observed dense packing of the adsorbing molecules. {While the overall charge of the protein vanishes at the pI, its surface still contains patches of positively and negatively charged amino acid residues. These charged patches drive the attraction to the negatively charged surface}\cite{Essa_2007}.\textsuperscript{,}\cite{Hartvig_2011} At pH values far from the isoelectric point, the proteins will repell each other and thus cause a less compact packing density on the adsorbing surface.

The validity of this model was examined experimentally\cite{pasche_2005} and theoretically\cite{Hartvig_2011} by studying the adsorption of lysozyme and $\alpha$-lactalbumin on differently charged surfaces. Taking into account the chemical properties of the ionizable groups and the orientation of the adsorbed protein with respect to the surface, the results of the experimental study were reproduced in a quantitative manner by just considering electrostatic interaction. Any other effects like dispersion forces, hydrophobic interactions, conformational changes or any other type of interaction were not required to reproduce the experimental findings. {One could thus assume that electrostatic interaction between charged patches, net charge and surface charges is the dominant mechanism behind protein adsorption}. Nevertheless, this interpretation is lacking at one point. The experiments employ surfaces at least partially covered with a tethered PEG spacer layer with a thickness exceeding the Debye length of the buffer. Contrary to the experimental reality, the theoretical description consideres an uncovered, blank surface. Thus the exact nature of the protein-surface interaction still remains unclear.

However, electrostatic interaction still seems to be exceptionally suitable to explain the behavior of folded \cytc on negatively charged silica\cite{qiao_2005}.

\subsection{Counterion Release Model}
\label{counterion_lit}

When a charged plate is immersed into a solution containing positive and negative ions the concentrations of these ions will change in the vicinity of the surface. Ions with the same sign as the surface charges (coions) will be repelled into the bulk while those with a different sign (counterions) will be drawn {towards} the surface. This leads to the formation of a shielding layer. For low ionic strength electrolytes and highly charged surfaces, the counterion concentration in the shielding layer strongly surpasses the bulk concentration. This leads to a considerable entropy loss for the bound ions.
When a second plate with opposite charge is brought in contact with the first plate, their charges mutually shield each other and the counterions are released into the bulk, since the shielding layer is no longer needed to ensure electroneutrality. This is accompanied by an entropy gain which causes an attractive force between the plates that adds to the mere coulombic attraction. Explicit calculations\cite{meier-koll_2004} of the resulting forces highlight the importance of this entropy-driven interaction especially for low ionic strength solutions. 

This model can be easily {extended} to protein adsorption, {at least in a qualitative manner}: Both the silica surface and the charged patches on the protein are covered with a shielding ion layer which desolves upon adsorption. For low ionic strength electrolytes, the entropic attraction of the counterion release should thus play an important part. Furthermore, if adsorption is dominated by counterion release, we can expect the binding affinity to drop significantly if the ionic strength increases.
This behavior has been observed experimentally for a multitude of different protein-surface combinations like \eg hemoglobin on clay\cite{causserand_2001}, \cytc on cyano-functionalized SBA-15\cite{Deere_2002}, \cytc on fused silica\cite{kraning_2007} and lysozyme in charged microgels\cite{welsch_2012}. 

Theoretical work confirms that this counterion release mechanism can indeed explain the experimentally observed Langmuir-type binding isotherms\cite{yigit_welsch}.

\section{Experimental}

\subsection{Cytochrome \textit{c}}
\label{materials_cytc}

Bovine heart cytochrome \textit{c} was purchased from Sigma Aldrich, catalog number C2037, and used as-{received} without further treatment or purification. It consists of a single polypeptide chain of 104 amino acid residues which are covalently bound via two cystein residues to a central heme complex. Its geometrical dimensions in the native, folded state have been reported\cite{Hartmann_2005} as $26\times 32\times 33 \Ang^3$.
Immersion into 8\,M urea solutions disrupts the hydrogen bonds which stabilize the \cytcol's tertiary structure and leads to unfolding of the molecule\cite{tsong_1975}. Small angle x-ray data\cite{Hsu_2007} of the unfolded cytochrome \textit{c} at pH\,7 {reveal} a structural transition from the almost spherical conformation of the folded protein (semi-major axis $18\,\Ang$, semi-minor axis $18\,\Ang$, radius of gyration $R_g=12.8\,\Ang$) to an eccentric ellipsoid shape (semi-major axis $65\,\Ang$, semi-minor axis $9\,\Ang$ at 8\,M urea and $R_g=29.7\,\Ang$ at 10\,M urea).

\subsection{Preparation of Mesoporous SBA-15 Silica}
\label{sba15-praep}

The synthesis of hexagonally ordered mesoporous SBA-15 was first reported by Zhao \etalol\cite{zhao1998} The samples used in this thesis were prepared according to the following {procedure}: We mix 4\,g of the tri-block co-polymer PEO$_{20}$-PPO$_{70}$-PEO$_{20}$ with 129.6\,g water and 19.3\,ml HCl (37\,\%). Due to its amphiphilic nature, the polymer forms an ordered phase of micellar structures when mixed with water. Vigorous stirring at 350 rpm for four hours is needed to ensure a homogenous emulsion.
The mixture is kept in an oil bath at 55\,\dC ~during this process. We then add 8.65\,g tetraethylorthosilicate {(TEOS)} and stir the system for another 20 hours. We subsequently increase the temperature to 85\,\dC ~and let the mixture rest for another 22 hours without stirring. During this time, the silicon from the TEOS leads to an accumulation of silica around the polymer micelles. These aggregates precipitate as a fine-grained powder.
Calcination of the repeatedly rinsed powder at 500\,\dC ~finally removes the polymer while preserving a negative of the micellar {structure} in the silica grains. The porous silica powder can now be used without further treatment or purification.
Small angle x-ray diffraction patterns were recorded at DESY, Hamburg. Five Bragg peaks were observed, confirming a hexagonal arrangement of linear mesopores with a lattice parameter of $a_h=10.71\pm0.08$~nm\cite{Hofmann2005, Zickler2006}.

\subsection{Nitrogen Sorption Isotherms}
\label{iso_append}

Nitrogen sorption isotherms were conducted by controlled filling (adsorption) and evacuation (desorption) of the sample via a custom-made gas handling system. The main part of the gas handling system was kept at room temperature while the sample cell was cooled to a well-known reference temperature. This was achieved by either using a closed-cycle helium refrigerator (Leybold RGD 510 Cryostat with RW 2 Compressor Unit) or by simply immersing the sample cell into a dewar vessel filled with liquid nitrogen. The pressure relaxation inside the system was measured using a Baratron Capacitance Manometer (MKS Intruments) with 1000\,torr full scale.

\subsection{Cytochrome \emph{c} Adsorption to SBA-15}
\label{experiment-ads}

Eppendorf Safe-Lock tubes were filled with small amounts of SBA-15. The mass of these SBA-15 samples (usually $4.0-9.0\,\tn{mg}$) was measured with an analytical balance (Sartorius type 1801). To provide for proper stirring during the experiment, we equipped each tube with a small stirring bar. Prior to the insertion the bars were rinsed with acetone and subsequently air-dried at 60\dC.
Cytochrome \emph{c} solutions were prepared with the following concentrations: 2000\,mg/l, 1500\,mg/l, 1250\,mg/l, 1000\,mg/l, 750\,mg/l, 500\,mg/l, 375\,mg/l, 250\,mg/l, 125\,mg/l and 42\,mg/l\label{desired_conc}. For each concentration two tubes were filled with 200\,$\mu$l of protein solution per 1\,mg of SBA-15. To ensure a fine dispersion of the silica grains, the samples were immersed into an ultrasonic bath for 10 minutes. To remove any impurities the solutions were filtered with disposable 0.8\,$\mu${m} syringe filters prior to the adsorption experiments.

During the adsorption process, the samples were kept in a heated water bath at 31\dC ~and stirred at 350 rpm using a IKA \emph{RCT basic safety control} magnetic stirrer. According to similar procedures reported in literature,\cite{Vinu_2004}\textsuperscript{,}\cite{zhang_2007} we let the system equilibrate for 5 days. Estimations of the adsorption kinetics indicate that this time is sufficient to reach equilibrium.

The amount of \cytc bound to the silica was calculated from the difference in the initial and final protein concentration.

\EPSB{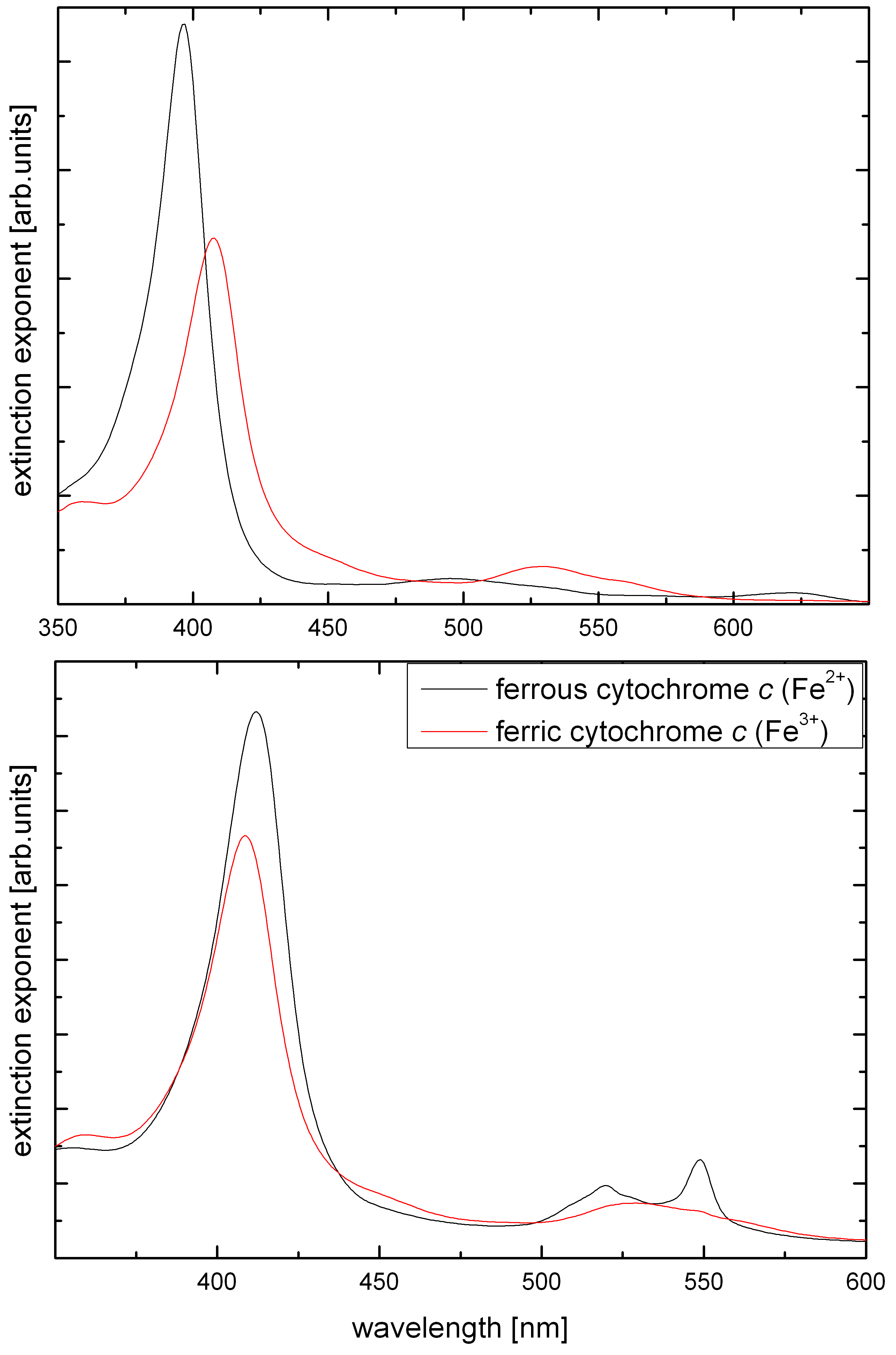}{.45\textwidth}{cytc_spectrum}{Top: Absorbance spectra of ferric \cytc in pure acidic buffer (red line) and in 8\,M urea acidic buffer (black line). Bottom: Spectra of ferrous (black) and ferric (red) \cytcol. Ferrous \cytc shows two distinct peaks at 520 and 550\,nm.}{Spectrum of folded and unfolded cytochrome \textit{c} under acidic conditions}

\subsection{Ionic Strength dependent Adsorption}

Protein solutions with a fixed concentration of 1\,g/l were prepared using buffers with an initial ionic strength of 10\,mM. We adjusted the ionic strength of each sample by mixing approriate amounts of pure and salinated buffers containing 1.11\,M NaCl. The addition of urea and salt altered the initial pH values of the buffers. However, these changes were found to be marginal and are therefore considered negligible. There is, however, one exception: The acidic buffer experiences quite substantial alterations of its pH value when salinated in the absence of urea. The implications of this changes will be discussed where necessary. Apart from the different preparations of the solutions, the ionic strength dependent measurements were carried out according to the same procedure as the adsorption isotherms mentioned above.

\subsection{Concentration Measurement via UV-Vis Spectroscopy}

The supernate concentration after the adsorption was determined photometrically with a \emph{Hitachi U-3501} spectrophotometer and disposable UV-Vis cuvettes (\emph{Plastibrand} PMMA 2.5\,ml macro cuvettes). Absorbance spectra of \cytcol, \ie the molar absorbtivity as a function of the wavelength $\alpha = \alpha (\lambda)$, are shown in figure \ref{cytc_spectrum}. The spectra are altered by chemical and structural transitions of the protein\cite{oellerich_2002}. We used the strong \emph{Soret band} absorption arround 410\,nm to simultaneously check the concentration and the conformational state of the protein. Additionally, we used the \emph{Q band} region between 500 and 600\,nm to monitor the oxidation state of the protein{: Samples containing reduced cytochrome \textit{c} show two distinct peaks in this region.} Since proper unfolding can only be achieved with oxidized \cytcol\cite{mclendon_1978}, any samples showing {signs} {of these peaks} were rejected from further evaluation. {In accordance with literature\cite{Hsu_2007}\textsuperscript{,}\cite{mclendon_1978}, complete unfolding of the oxidized protein was assumed in urea solutions without further experimental confirmation.}

\subsection{Buffer Solutions}

All buffers were prepared in-house according to the following recipes. PH-dependent measurements used buffers with pH 3.0 (0.018\,M citric acid and 0.033\,M trisodium citrate), pH 6.0 (0.025\,M monopotassium phospate and 0.002\,M sodium hydroxide), pH 9.0 (0.018\,M glycin, 0.017\,M sodium chloride and 0.03\,M sodium hydroxide). 
The high urea content needed to unfold the proteins considerably altered the pH value of the buffers to 4.42, 6.40 and 9.69, respectively. To ensure comparability between the measurements of folded and unfolded \cytcol, new buffers were prepared for the measurements without added urea as follows: pH 4.4 (0.01\,M trisodium citrate and 0.011\,M citric acid) and pH 9.7 (0.0135\,M glycin, 0.014\,M sodium chloride and 0.073\,M sodium hydroxide). The pH change in the near-neutral buffer was found small enough to use the same buffer for the measurements with and without urea since both the initial and the altered pH were comfortably in the region were protein and surface bear opposite charge signs.
Ionic strength dependent measurements used the following buffers: pH 3.0 (116.37\,g 0.01\,M citric acid and 4.0\,ml 0.01\,M trisodium citrate), pH 3.8 (200.02\,g 0.01\,M acetic acid and 15\,ml 0.01\,M sodium acetate), pH 4.5 (100.23\,g 0.01\,M trisodium citrate and 120\,ml 0.01\,M citric acid), pH 6 (100.14\,g 0.01\,M monopotassium phospate and 32.5\,ml 0.01\,M sodium hydroxide), pH 7.3 (50.40\,g 0.01\,M sodium hydroxide and 41.7\,ml 0.05\,M monopotassium phospate), pH 8.5 (83\,ml 0.01\,M sodium tetraborate and 13.1\,ml 0.01\,M hydrochloric acid) and pH 10.6 (100\,g 0.01\,M sodium bicarbonate and 71\,ml 0.01\,M sodium hydroxide).

\section{Results}
\subsection{SBA-15 Characterization}

\EPSB{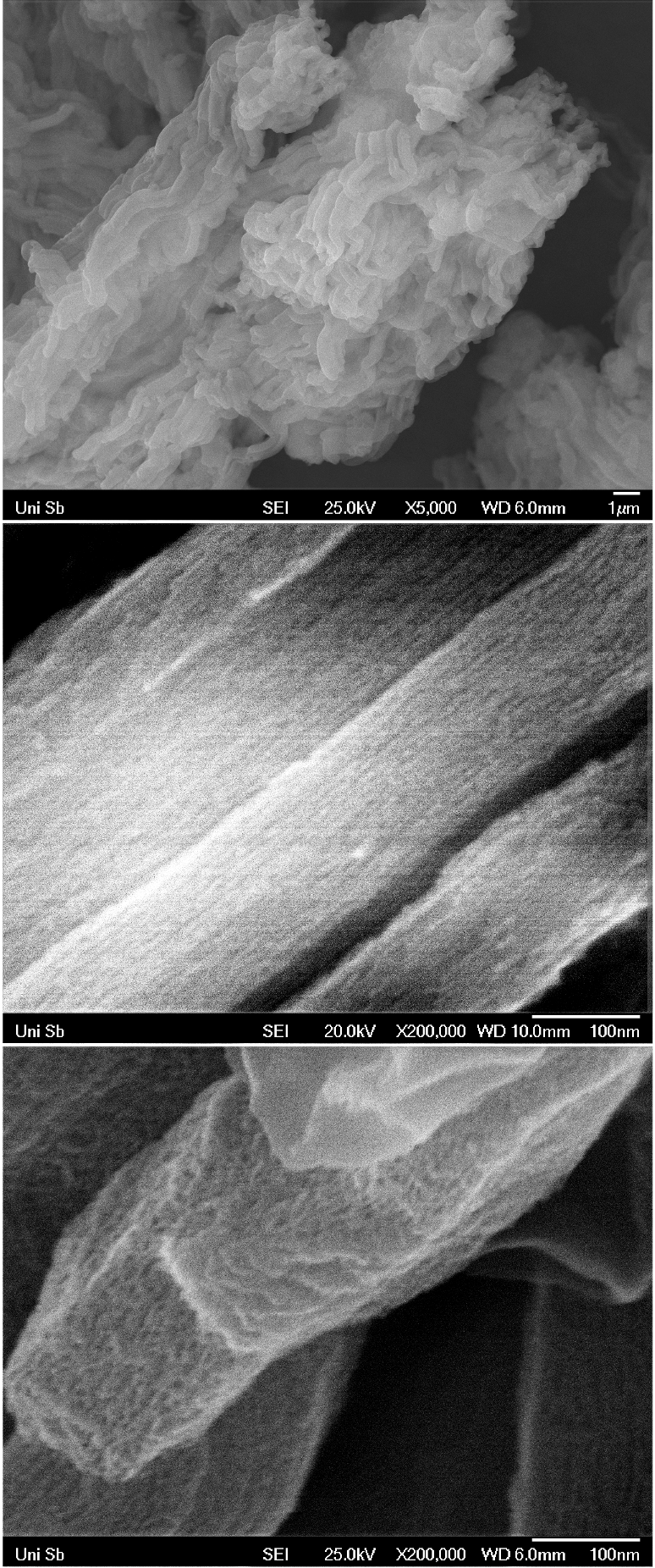}{.35\textwidth}{sem-sba15}{Scanning electron micrograph of SBA-15 powder. The contrast of all pictures was enhanced with standard image processing software (GNU Image Manipulation Program GIMP 2.6.5). Top: Aggregate of individual grains. Middle: Lateral view of a single grain. Bottom: The front edge of a silica grain.}{Scanning electron micrographs of SBA-15}

Figure \ref{sem-sba15} shows scanning electron micrographs of the SBA-15 powder used in this {study}. The powder consists of needle-like aggregates of micrometer-sized silica grains. A zoom in on the edge of such a single grain reveals the porous structure. The picture in the bottom panel shows a micrograph of the front edge of a grain. The pore openings are visible as dark spots on the brighter silica surface. The middle panel presents a lateral view of the grain surface where the pores can be identified as dark grooves. Their high aspect ratio is evident in this micrograph.

Nitrogen sorption measurements were performed as outlined above. A sorption isotherm is the plot of the total amount of adsorbed nitrogen $n=\sum \Delta n_{samp}$ versus the {corresponding} relaxation pressure $p_{relax}$. For the sake of simplicity and comparability, we usually use the dimensionless quantities filling fraction $f=n/n_0$ and reduced vapor pressure $P=p/p_0$. Here, $n_0$ is the amount of nitrogen needed for a complete filling of the sample's micro- and mesoporous structures. As can be seen in the sorption isotherm shown in figure \ref{iso-scfit}, there is some condensation beyond $n_0$. This can be attributed to highly irregular macroporous structures. $p_0$ denotes the bulk vapor pressure at the sample temperature.

\EPSB{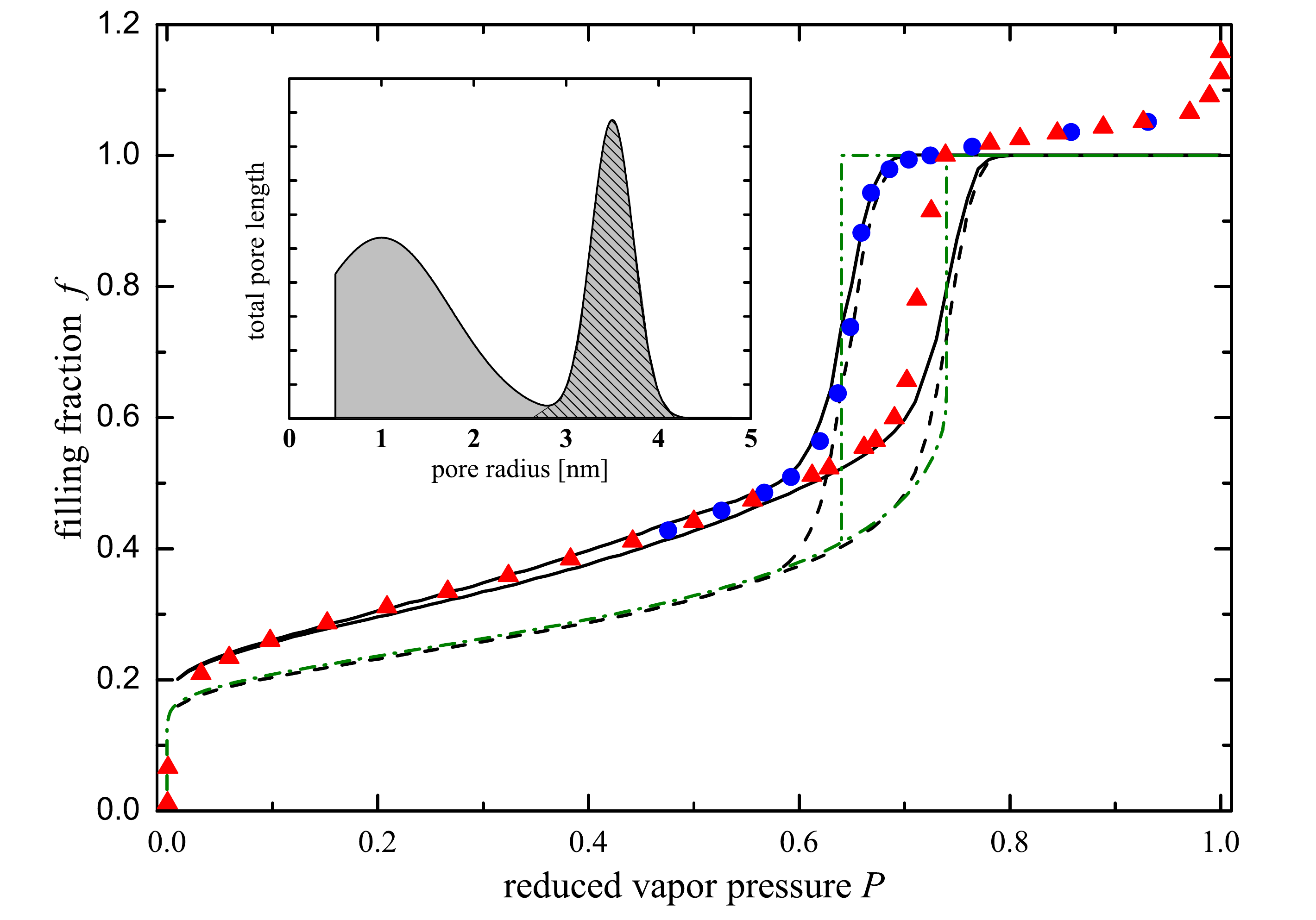}{.45\textwidth}{iso-scfit}{Experimental nitrogen adsorption (triangles) and desorption (circles) isotherm and theoretical isotherms according to the Saam and Cole Theory for monodisperse pores (dash-dotted line), Gaussian pore size distribution (dashed line) and the bimodal pore size distribution (solid line) shown in the inset. This figure is taken from {a previous study} \cite{moerz_2012}.}{Evaluation of an experimental isotherm with the SC theory}

At very low pressures the interaction between the gas molecules and the naked surface causes the gas to condensate. The large initial slope of the isotherm reflects this phenomenon. The kink in the isotherm at a reduced pressure of approximately 0.05 marks the point were a complete monolayer has formed. Further increasing of the pressure leads to a growth of the condensed layer's thickness. This region which spans up to reduced pressures of 0.6 is called the reversible multilayer regime. Upon reaching a critical thickness the liquid film becomes unstable. Further adsorption of gas molecules then leads to the spontaneous formation of \emph{capillary bridges}, \ie small liquid droplets which span across the entire pore diameter. In figure \ref{iso-scfit} this transition can be seen for p/p$_0$ between 0.6 and 0.7 as a strong increase in the isotherm's slope. This condensation is a first order phase transition. Adding more gas molecules to the system will cause the droplets to grow longitudinal to the pore axis while the pressure remains unaltered. Note that there is a strong hysteresis in the pressure associated with capillary condensation between adsorption and desorption.

We analyze the recorded isotherms using the mean-field model by Saam and Cole\cite{sc1974}.\textsuperscript{,}\cite{sc1975} Contrary to much simpler models like the BET or the BJH approach, this theory explicitely includes the van der Waals interaction between condensate and pore wall.

The green dash-dotted line in figure \ref{iso-scfit} corresponds to the hypothetical isotherm of a sample with a monodisperse mesopore radius of $r_p=3.3\,$nm, while the black dashed curve includes a Gaussian distribution of pore sizes. We calculated 150 individual Saam and Cole isotherms, each with a different pore radius but otherwise identical parameters. The final isotherm is a sum of these isotherms, weighted according to a single Gaussian distribution and normalized to a filling fraction $f=1$ at $p/p_0=1$. It reproduces the experimental hysteresis loop rather well, but severely underestimates the multilayer region. This can be fixed by using a bimodal pore size distribution using two Gaussian distributions. Refering to the SAXS data by Impéror-Clerc \etalol\cite{imperorclerc2000}, we attribute those two peaks to the hexagonal mesopore array and a microporous layer surrounding the individual pores. We also added a lower cut-off of 0.5\,nm to the pore sizes allowed in the calculation. This cut-off was chosen for numerical reasons, but it is also reasonable to assume a physical cut-off for the SBA-15's microporosity: Since the micropores presumably stem from individual hydrophilic polymer segments, no micropore should be smaller than the backbone of the polymer chains. The samples used in this study show a broad peak at $r_{micro}=0.75$\,nm with $\sigma_{micro}=78\%$ and a narrower mesopore peak at $r_{meso}=3.3$\,nm with $\sigma_{meso}=6.5\%$.

\subsection{PH-Dependent Cytochrome \textit{c} Adsorption}
\label{results_ads}

The results of the protein adsorption experiments are shown in figure \ref{cytc_ads}. The experiments were done at three different pH values corresponding to three fundamentally disparate electrochemical conditions. The red line and symbols correspond to a pH of 9.7, close to the isoelectric point of the \cytcol. The protein is virtually uncharged under these conditions, while the silica exhibits strong negative surface charges. The green line and symbols were measured under near-neutral conditions where both the protein and the surface are charged with opposite signs, resulting in mutual attraction. Finally, the blue lines and symbols represent the measurements near the SBA-15's isoelectric point. The protein has a strong positive overall charge at this pH, while the surface is mostly neutral (The isoelectric point of SBA-15 has been reported as pH\,3.8\cite{Essa_2007}.).
The solid lines in the figure are Langmuir-type fits defined by

\EQ{langmuir_iso}{n_{ads} (c_s) = \frac{n_L \cdot \alpha \cdot c_s}{1 + \alpha c_s}}{.}

This equation relates the amount of adsorbed protein $n_{ads}$ to the residual supernate concentration $c_s$, using the amount of protein $n_L$ needed to form a complete monolayer covering the sample surface and an interaction parameter $\alpha$ as fit parameters. This behavior is well-known: Isotherms similar to the ones presented here were already presented by Vinu \etalol.\cite{Vinu_2004}.

\EPSB{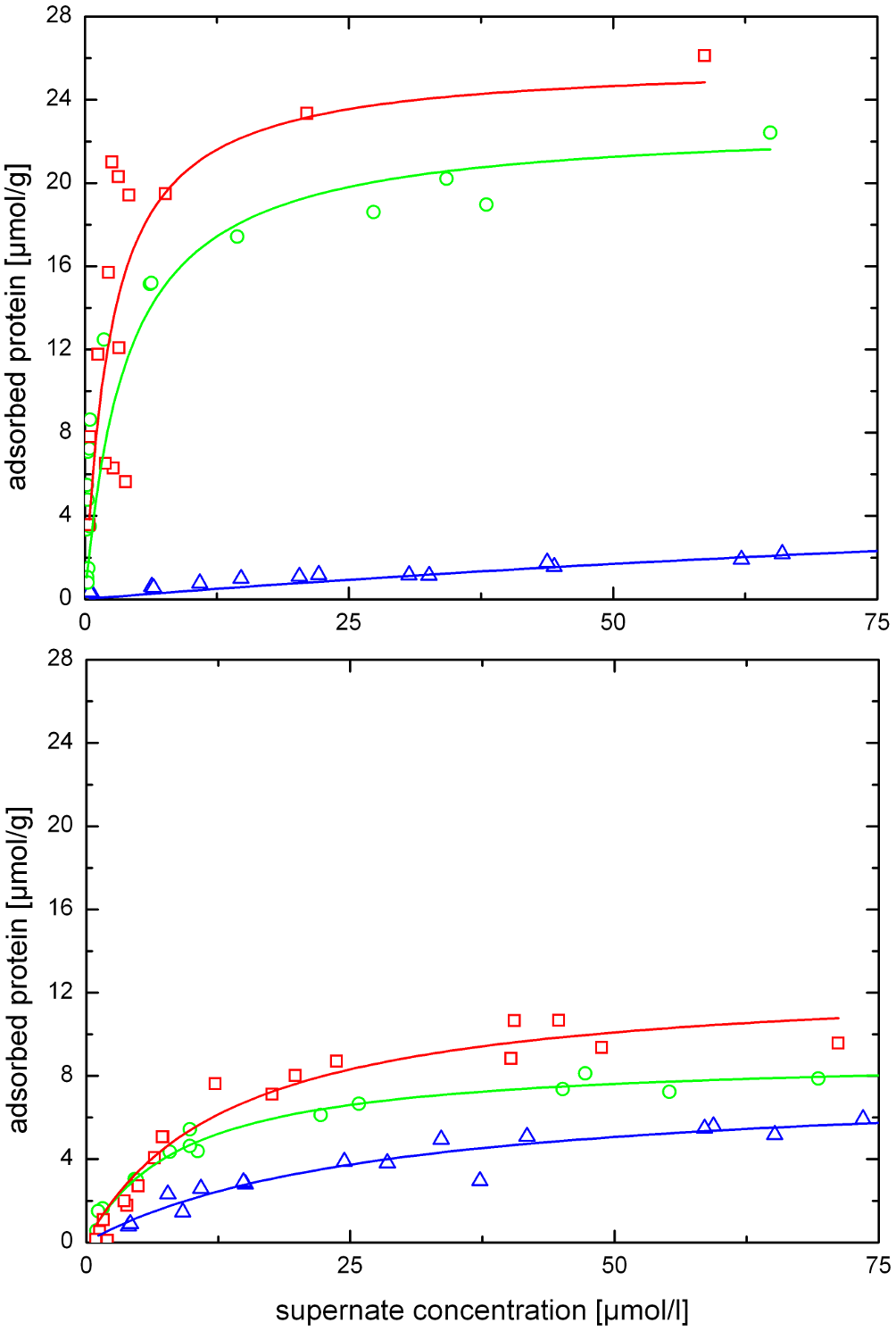}{.45\textwidth}{cytc_ads}{Cytochrome \emph{c} adsorption isotherms for pure buffer (top) and for 8\,M urea buffer (bottom). Solid lines are Langmuir-type fits to the experimental data. The fit parameters are listed in table \ref{tab_langmuir}. The isotherms were recorded pH\,9.7 (red squares), pH\,6.4 (green circles) and pH\,4.4 (blue triangles).}{Langmuir isotherms of cytochrome \textit{c} in SBA-15} 

For both folding states of \cytc the pore loadings, \ie the amount of adsorbed protein, decrease with decreasing pH value. This decrease is quite moderate for unfolded \cytc but severe for the native, folded type. While the native \cytc adsorbs up to 25.9$\pm$4.3\,$\mu$mol per gram of SBA-15 near its pI and still up to {22.9}$\pm${7.4}\,$\mu$mol/g at pH 6.4, the adsorption is almost negligible on the weakly charged surface at pH\,4.4 with a mere 3.4$\pm$0.4\,$\mu$mol/g saturation value and {an} interaction parameter $\alpha=0.023\pm0.006$\,l/$\mu$mol, almost 20 times smaller as for the alkaline buffer. Unfolded \cytcol, on the other hand, seem less susceptible to changes in the buffer pH. Both the saturation value of the pore loading and the interaction parameter at pH\,4.4 are still about half of the corresponding values at the protein's pI.

To evaluate what percentage of the available pore space is actually occupied by adsorbed molecules, the volume and surface of the mesopores were extracted from the nitrogen sorption data and the calculated pore size distribution. The samples have a mesopore volume of $V_{meso}=0.520\,\tn{cm}^3/\tn{g}$ and a specific surface area $A_{meso}=317\,\tn{m}^2/\tn{g}$.
The geometrical dimensions\cite{Hartmann_2005} of folded \cytc ($26 \times 32 \times 33\,\tn{\AA}^3$) yield a volume of $V_{f}=14.4\,\tn{nm}^3$ per molecule. Assuming a hexagonal packing of the individual molecules in a surface-covering monolayer, each protein takes up $11.2\,\tn{nm}^2$ of the surface area. A complete and perfect monolayer would thus correspond to $1.48\cdot10^{-11} \, \tn{mol}/\tn{cm}^2$, which is in accordance with the literature\cite{cheng_2003}. With these data, the highest pore loading of our experiments (25.9\,$\mu$mol/g) implies a packing density of 8.1\,$\cdot10^{-12} \, \tn{mol}/\tn{cm}^2$ taking up 43\% of the mesopore volume. This is only about half of the expected value, but since the highly curved and probably corrugated pore surface is unlikely to allow a perfect hexagonal protein assembly this deviation is within reasonable limits. {We also expect the small size of  the pores to impose a steric hindrance to perfect monolayer adsorption.} It is further possible that the protein experiences conformational changes upon adsorption which increase the area that is occupied by each single molecule. Our value for the percentage of the pore volume which is filled upon adsorption coincides well with previous findings by Miyahara \etalol\cite{miyahara_2006}. They found that adsorption of \cytc takes up approximately half the pore volume of SBA-15. Given the respective radii of the protein and the pores, they calculate that a close packing of spherical molecules inside a cylinder takes up 48\% of the available cylinder volume. We therefore assume that in the case of adsorption of the native protein at pH\,9.7 the pores of our sample are completely stuffed with \cytcol. At pH\,6.4 the surface and volume packing densities are {7.1}\,$\cdot10^{-12} \, \tn{mol}/\tn{cm}^2$ and {38}\%, respectively. At pH\,4.4, we observe a surface coverage of 1.06\,$\cdot10^{-12} \, \tn{mol}/\tn{cm}^2$ and a volume packing density of 5.6\%.

We obtain the respective values for the unfolded protein using the x-ray scattering data of Hsu \etalol\cite{Hsu_2007}. In 8\,M urea solutions, each molecule has a volume of $V_{u}=22\,\tn{nm}^3$ and takes up $23.4\,\tn{nm}^2$ when adsorbed on a surface. The theoretical monolayer would thus contain $7\cdot10^{-12} \, \tn{mol}/\tn{cm}^2$. Again, the experimental data are about half of this value. At pH\,9.7 we observe a coverage 4.0\,$\cdot10^{-12} \, \tn{mol}/\tn{cm}^2$ and a packing density of 33\%. At pH\,6.4 we observe 2.8\,$\cdot10^{-12} \, \tn{mol}/\tn{cm}^2$ and 23\% packing density while pH\,4.4 yields 2.5\,$\cdot10^{-12} \, \tn{mol}/\tn{cm}^2$ and 20\%, respectively. 

\begin{table*}[h]
	\centering
	\caption{Fit parameters $n_L$ and $\alpha$ of the Langmuir-type isotherms in figure \ref{cytc_ads} and their respective margins of error. Folded samples are marked as \emph{f}, unfolded as \emph{u}.}
	\label{tab_langmuir}
	\begin{tabular}{|l|c|c|c|c|}
	\hline 
	& $n_L$ [$\mu$mol/g] & $\Delta n_L$ [$\mu$mol/g] & $\alpha$ [l/$\mu$mol] & $\Delta$$\alpha$ [l/$\mu$mol] \\
 	\hline
pH 4.4 \emph{f} & 3.37 &  $\pm$ 0.40 & 0.023 & $\pm$ 0.006  \\
pH 4.4 \emph{u} & 7.84 & $\pm$ 0.56 & 0.036 & $\pm$ 0.007  \\
pH 6.4 \emph{f} & 22.9 & $\pm$ 7.4 & 0.25 & $\pm$ 0.10  \\
pH 6.4 \emph{u} & 9.00 &  $\pm$ 0.22 & 0.11 & $\pm$ 0.01  \\
pH 9.7 \emph{f} & 25.9 & $\pm$ 4.3 & 0.40 & $\pm$ 0.19  \\
pH 9.7 \emph{u} & 12.8 & $\pm$ 0.9 & 0.073 & $\pm$ 0.014  \\

	\hline
	\end{tabular}
	
\end{table*}

\subsection{Influence of the Ionic Strength}

Figure \ref{ionic_strength} depicts the adsorbed amount of \cytc as a function of the ionic strength. The data refer to the actual number of proteins adsorbed from an 1\,g/l solution and must not be confused with the monolayer coverage $n_L$ from the Langmuir-type fits. However, 1\,g/l equals an initial concentration of approximately 80\,$\mu$mol/l. In the limit of low adsorption where the initial and residual supernate concentration differ only slightly this should be sufficient to ensure that the Langmuir isotherm is almost saturated. Apart from the folded \cytc samples in alkaline buffer, it is therefore not too far-fetched to use the measured adsorbed amount as an estimate of $n_L$. The pH values differ slightly from the ones used in the previous experiments but still correspond to the two distinct isoelectric points and the near-neutral conditions. Black symbols show the measurements with pH\,10 buffer, red symbols used pH\,6.5 buffer and the blue symbols correspond to pH\,3.8 buffer. The folded protein data for the alkaline buffer exhibit a maximum at an ionic strength 0.2\,M/l and a decrease to approximately 10\,$\mu$mol/g at 1\,M/l NaCl. The data from the pH\,6.5 buffer do not show this maximum but also a monotonic decrease to 7\,$\mu$mol/g at 0.4\,M/l and become only weakly susceptible to further changes in the salinity. The measurements using the acidic buffer showed a quite different behavior. Starting at roughly 2\,$\mu$mol/g at low salt content, the adsorbed amout of protein increases almost linearly to 5\,$\mu$mol/g at 1\,M/l.
The adsorbtion of unfolded \cytc is much more strongly influenced by salinity. Adsorption from alkaline buffers decreases strongly until saturating at about 2.5\,$\mu$mol/g above 0.7\,M/l NaCl. Adsorption from near-neutral buffers even becomes almost negligible for 0.3\,M/l and higher. Again, the acidic buffers show a different behavior. Both for very high and low salinities, the adsorption ranges around 2.0-2.5\,$\mu$mol/g, but undergoes a minimum of approximately 1\,$\mu$mol/g at 0.4\,M/l.

\EPSB{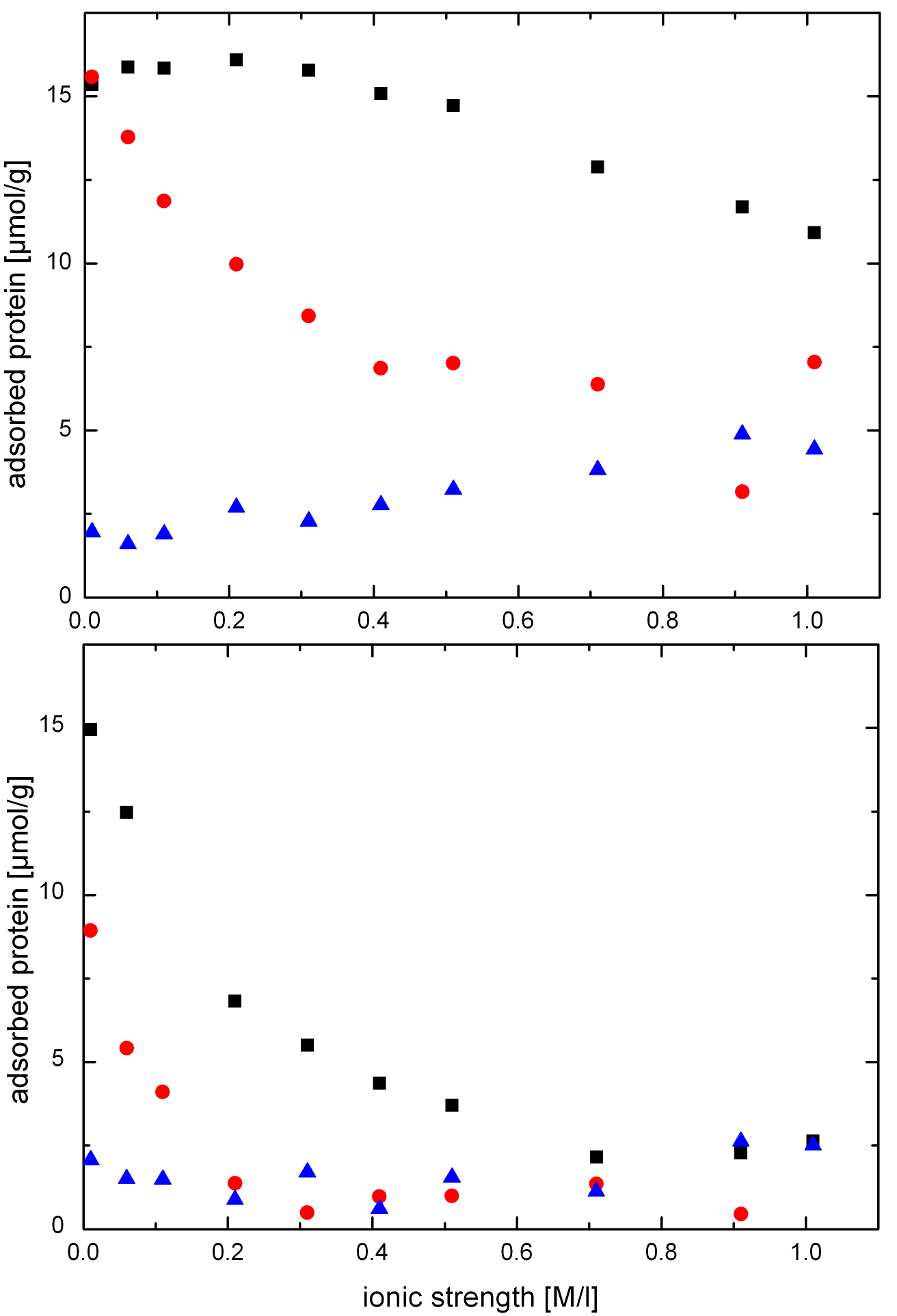}{.5\textwidth}{ionic_strength}{Equilibrium value of \cytc adsorption from a 1\,g/l solution as a function of ionic strength for pure buffer (top) and for 8\,M urea buffer (bottom) at pH\,10.0 (black squares), pH\,6.5 (red circles) and pH\,3.8 (blue triangles).}{Maximum \cytc pore loading as function of ionic strength} 

\section{Discussion}
\subsection{Electrostatic Interpretation}

As expected from previous studies,\cite{Vinu_2004}\textsuperscript{,}\cite{Hartmann_2005} the adsorption of folded \cytc can be readily explained with the electrostatic interaction model. The highest pore loading is found at the isoelectric point of \cytcol, where the protein-protein repulsion vanishes and the protein-wall attraction facilitates high packing densities. At intermediate pH, the packing density decreases due to the rise of electrostatic protein-protein repulsion. Finally, at the silica's pI, protein adsorption is largely supressed due to vanishing silica surface charges. This is not only reflected in the very small saturation value of the Langmuir isotherm $n_{0, pH=4.4}=3.37\,\mu$mol/g, but also in the severe decrease of the interaction parameter $\alpha$, which drops by almost a factor of 20 from $0.40\pm0.19$\,l/$\mu$mol at pH\,9.7 to $0.023\pm0.006$\,l/$\mu$mol at pH\,4.4.

This picture also applies to the adsorption of the unfolded protein. Again, the highest pore loading is observed at the protein's pI, while the binding affinity is rather low at pH\,4.4. The maximum pore loadings of unfolded protein are approximately half the value of the folded type. This fits well with the surface area occupied by a single molecule, which increases from 11.2\,nm$^2$ to 23.4\,nm$^2$ upon unfolding.

There is, however, one puzzling point: The isotherm of unfolded \cytc in the acidic buffer does not experience the severe drop in both its parameters compared to the folded protein. Electrostatic interaction alone fails to explain this behavior. We found it plausible to assume that van der Waals forces cause this different behavior. At the silica's pI they probably dominate the protein binding due to the absence of electrostatic interaction. The folded protein has a relatively rigid shape. While some data suggests small conformational changes upon adsorption\cite{menaa_2008}, these changes are rather minor. As a consequence, the folded protein has a rather small contact area with the silica and can not exploit these interactions very effectively. The unfolded protein, on the other hand, has a much higher contact area with the silica and is probably rather flexible\cite{kawahara_1966_2}. It should therefore be much more capable to exploit the van der Waals interaction and still be able to adsorb to a relatively high extend even in the absence of electrostatic attraction.

To test this hypothesis, we examined the influence of the buffer's salinity on the protein adsorption. If it is true that the unfolded protein's adsorption behavior is influenced by van der Waals forces and the folded protein's behavior is solely due to {electrostatic} interaction, changes in the salinity should affect the unfolded samples less than the folded ones. A single glance at figure \ref{ionic_strength} reveals that the opposite is true: The adsorption of unfolded \cytc is highly susceptible to changes in the ionic strength and becomes almost {negligible} above 0.7\,M/l NaCl. Folded \cytcol, on the other hand, adsorbs still quite well even at high salinity, especially at the protein's pI.

{Interfacial charge regulation has to be taken into account when discussing electrostatically driven adsorption. As discribed in detail by Hartvig \etalol\cite{Hartvig_2011}, the enrichment of counterions over a charged surface leads to a alteration of the pH-value in the Debye layer. This altered pH-value will affect the local charges on the protein and might lead to a behavior which differs from what would be expected from the bulk solution's properties. However, a quantitative inclusion of this mechanism into our interpretation encounters two problems. First, the calculation of the surface pH requires knowledge of the surface potential, which is influenced not only by the protonation of the silica's silanol groups, but also on the amount and charge of the adsorbed proteins. Second, even for a known alteration of the surface pH the impact on the protein adsorption can only be calculated with precise knowledge of the pH-dependent charge profile of the protein, its structure, the spatial distribution of charged surface patches and the orientation of the molecule upon adsorption. To the best of our knowledge, these data do not exist for unfolded cytochrome c. We can, however, present some qualitative considerations. At the acidic pH, the charge of the silica surface vanishes and thus charge regulation should not play a major role. At higher pH, the adsorption of the (presumably positively charged) proteins will screen the negative surface charge to a certain extend and suppress charge regulation. The fact that the Langmuir isotherms for cytochrome \textit{c} presented here and in literature can be readily explained without assuming charge regulation suggest the it plays only a negligible role in the model system used here.}

\subsection{Counterion Release Interpretation}

The surface of a protein is composed of positively and negatively charged patches of different size as well as polar and hydrophobic regions\cite{becker_welsch_2012}. The counterion release mechanism can be used to explain protein adsorption to charged surfaces simply by considering the local shielding of the charged surface patches. Becker \etalol\cite{becker_welsch_2012} argue that counterion release is an effective mechanism as long as the ionic strength of the bulk solution is smaller than the surface charge concentration of the charged patches. As an example they mention $\beta$-lactoglobulin which has 5 charges in a patch of 10\,$\tn{nm}^2$, corresponding to an ion concentration of 0.3\,M, assuming that the shielding ions are confined within one Debye length from the surface. Thus, any attraction from counterion release should vanish for higher ionic strength. Of course the exact values might be different for \cytcol, but without precise knowledge of the cytochrome's surface charges at the various chemical conditions examined we might still use the given example as a reasonable, semi-quantitative reference.

With this value in mind, we can interpret the data from figure \ref{ionic_strength}. If we neglect the data for the acidic buffers we notice the general trend of decreasing adsorption for increasing salinity. This decrease seems not very strong for folded protein in the alkaline buffer. But note that the starting concentrations were too low to saturate the adsorption, so the data might underestimate the true extent of this decrease. Adsorption of folded \cytc at pH\,6.5 strongly decreases for {increasing} ionic strength and seems to saturate above 0.4\,M. \label{blubb}This coincides well with the estimated ion concentration limit of Becker \etalol. Any adsorption at higher ionic strength is likely due to van der Waals forces. Unfolded \cytc behaves quite similarly: The curve for pH\,10 saturates somewhere between 0.5\,M and 0.7\,M while adsorption is merely over the detection limit for pH\,6.5 above 0.2\,M. These data indicate that counterion release is indeed the driving mechanism behind \cytc adsorption.

However, there are still some features which the counterion release model fails to account for. The high pore loadings of folded \cytc at pH\,10 can only be explained if we assume highly charged, small surface patches. The increase of the pore loading for both conformational states in acidic buffers above 0.4\,M also remains unexplained. Concerning the data from figure \ref{cytc_ads}, the absence of the sudden drop in unfolded protein pore loading at pH\,4.4 is still unaccounted for and the higher pore loadings at higher pH values can only be explained if we make the \emph{ad hoc} assumption that the charge density of the patches increases upon approaching the isoelectric point. Not only would this assumption be overly tentative, it is also highly counterintuitive given the fact that the overall charge of the protein vanishes at the pI. We therefore conclude that the counterion release mechanism alone is also insufficient to explain \cytc adsorption.

\subsection{Complete Interpretation}

None of the simple models presented above is sufficient to exclusively account for the observed protein adsorption behavior. The driving force behind \cytc immobilization on silica is probably a rather complex compilation of several interactions. We propose a comprehensive explanation of the experimental data using electrostatic and counterion release interaction as well as van der Waals forces.

The electrostatic interaction can account for the high pore loadings close to the proteins isoelectric point, which are rather puzzling if we concern only the counterion release. It furthermore explains the small maximum of the ionic strength dependent pore loading for folded \cytc at pH\,10. Even though the adsorption is unsaturated for these samples, the maximum at 0.2\,M NaCl is unambiguous and indicates a higher binding affinity. While residual interprotein repulsion will probably play a role at very low ionic strength it is likely shielded at 0.2\,M, facilitating a denser packing.
Both the counterion release and the electrostatic interaction will suffer from increasing ionic strength. The first will do so due to a decreasing entropy gain while the latter will simply be shielded by the solution's high ion concentration and thus shorter Debye length. This behavior is very well reflected in the ionic strength dependent measurements of figure \ref{ionic_strength}. The estimation of Becker \etalol\cite{becker_welsch_2012} that the counterion release mechanism should be ineffective at salinities above 0.3\,M coincides well with our data, as outlined above. It is still not entirely clear why the unfolded protein is more susceptible to salinity changes. {Without precise knowledge of the unfolded protein's structure and orientation, we can not rule out interfacial charge regulation as a possible explanation. But given the considerartions outlined above, we propose an interpretation using the counterion release mechanism: S}ince the unfolded molecules have a much higher surface area it is not very far-fetched to assume that their charged amino acid residues are distributed over a larger area. This would reduce the ion concentration needed to shield the patches and reduce the ionic strength at which the counterion release becomes effectless. This reduced effectiveness of the counterion release might also be reflected in the Langmuir interaction parameters $\alpha$. With the exception of the native sample at pH\,4.4, the values for $\alpha$ at a given pH are considerably smaller for the unfolded samples. Since $\alpha$ represents the interaction strength, this is consistent with the assumption of a lowered charge density in the unfolded state.

At the silica's isoelectric point both electrostatic and counterion release interaction will cease to work. Any residual adsorption will therefore be dominated by van der Waals forces. Unfolded \cytc has a higher surface area and is probably more flexible than its native counterpart. This enables the protein to squeeze tightly to the surface and exploit the van der Waals interaction rather effectively. It can therefore still adsorb quite well in the acidic buffers whereas the folded protein exhibits a very low binding affinity.

The peculiar behavior of the ionic strength-dependent measurements for acidic buffers is likely an artefact of the buffers used. While the pure buffer with pH\,3.86 is very close to the silica's isoelectric point, adding 1\,M NaCl severely reduces the pH to 3.2. Thus, increasing salinity will lower the buffer pH below the silica's pI. The pore surface will exhibit positive charges and allow for electrostatic binding with negative patches on the protein, leading to the almost linear pore loading increase oberserved for the blue data points in the upper panel of figure \ref{ionic_strength}. Additionally, higher salinities will enhance the pore loading by shielding the interprotein repulsion. This behavior is slightly altered when urea is added. The pure buffer with urea has a pH of 3.96. At this pH, the silica is still slightly charged, allowing for some residual electrostatic binding. Adding 1\,M NaCl alters the pH to 3.7, which is below the silica's pI. Again, a small amount of surface charges will be present, although with different sign. At some salinity in between, the pI will be met exactly, resulting in almost negligible binding. 

A last aspect which remains unaccounted for is the high pore loading of folded \cytc at high salinity close to its pI. The pore loading is still higher than 10\,$\mu$mol/g, about twice the value observed for the folded protein in the other buffers and about four times that of the unfolded samples. Interprotein repulsion should be shielded anyway at this ionic strength, so the difference between the alkaline and near-neutral measurements can not be explained within the scope of electrostatic interaction. Van der Waals forces are not accountable for this, either, since they are independent of the chemical conditions. And without precise knowledge of the surface charge distribution under these very specific chemical conditions it is impossible to rule out {contributions from counterion release or charge regulation. Note that we assume that the protein structure is identical for pH\,6.4 and pH\,9.7. However, cytochrome \textit{c} undergoes the alkaline transition at the latter pH. This transition is marked by the replacement of the Met-80 ligand of the heme group by a lysine and results in a structure which is still folded but more flexible than the native one\cite{Perroud2005}. Analogous to the explanation used for unfolded cytochrome \textit{c} this enhanced flexibility might cause higher pore loadings.}


\section{Conclusions}

The adsorption of bovine heart \cytc to the mesoporous silica powder SBA-15 has been studied for different folding states of the protein and for a set of different chemical conditions. The adsorption of folded \cytc is usually explained in literature by a simple electrostatic interaction model where the molecules bind to the charged silica via charged amino acid residues on the protein's surface. The packing density is mainly defined by the repulsion between the individual protein molecules which is defined by their overall net charge. By trying to extend this model to the adsorption behavior of unfolded \cytcol, we find that this simple model fails to account for the complicated dependence of protein binding on both the ionic strength and the pH of the buffer solution. We tried a different approach using the counterion release mechanism which readily explains the differences between the two folding states. However, the counterion release can not explain the residual adsorption observed for high ionic strength and the high pore loading at the protein's isoelectric point. 
Only by combining electrostatic interaction, counterion release and van der Waals forces, we are able to explain the binding behavior of \cytc to silica. Apparently, protein binding to charged inorganic surfaces is caused by a complex interplay of different mechanisms.
The application of these findings for the fractionation of mixed protein solutions will be part of a subsequent {study}.\\

\begin{acknowledgement}

This work has been supported by the graduate school 1276 of the German Research Foundation (DFG), `Structure formation and transport in complex systems' (Saarbr\"{u}cken).

\end{acknowledgement}



\begin{mcitethebibliography}{41}
\providecommand*\natexlab[1]{#1}
\providecommand*\mciteSetBstSublistMode[1]{}
\providecommand*\mciteSetBstMaxWidthForm[2]{}
\providecommand*\mciteBstWouldAddEndPuncttrue
  {\def\EndOfBibitem{\unskip.}}
\providecommand*\mciteBstWouldAddEndPunctfalse
  {\let\EndOfBibitem\relax}
\providecommand*\mciteSetBstMidEndSepPunct[3]{}
\providecommand*\mciteSetBstSublistLabelBeginEnd[3]{}
\providecommand*\EndOfBibitem{}
\mciteSetBstSublistMode{f}
\mciteSetBstMaxWidthForm{subitem}{(\alph{mcitesubitemcount})}
\mciteSetBstSublistLabelBeginEnd
  {\mcitemaxwidthsubitemform\space}
  {\relax}
  {\relax}

\bibitem[Washmon-Kriel et~al.(2000)Washmon-Kriel, Jimenez, and
  Balkus]{washmon-kriel_1999}
Washmon-Kriel,~L.; Jimenez,~V.~L.; Balkus,~K.~J. Cytochrome \textit{c}
  immobilization into mesoporous molecular sieves. \emph{J. Mol. Catal. B:
  Enzym.} \textbf{2000}, \emph{10}, 453--469\relax
\mciteBstWouldAddEndPuncttrue
\mciteSetBstMidEndSepPunct{\mcitedefaultmidpunct}
{\mcitedefaultendpunct}{\mcitedefaultseppunct}\relax
\EndOfBibitem
\bibitem[Quinn et~al.(2008)Quinn, Mantz, Jacobs, Bellion, and
  Santen]{Quinn2008}
Quinn,~A.; Mantz,~H.; Jacobs,~K.; Bellion,~M.; Santen,~L. Protein adsorption
  kinetics in different surface potentials. \emph{Epl} \textbf{2008},
  \emph{81}, 56003\relax
\mciteBstWouldAddEndPuncttrue
\mciteSetBstMidEndSepPunct{\mcitedefaultmidpunct}
{\mcitedefaultendpunct}{\mcitedefaultseppunct}\relax
\EndOfBibitem
\bibitem[Evers et~al.(2011)Evers, Steitz, Tolan, and Czeslik]{Evers2011}
Evers,~F.; Steitz,~R.; Tolan,~M.; Czeslik,~C. Reduced Protein Adsorption by
  Osmolytes. \emph{Langmuir} \textbf{2011}, \emph{27}, 6995--7001\relax
\mciteBstWouldAddEndPuncttrue
\mciteSetBstMidEndSepPunct{\mcitedefaultmidpunct}
{\mcitedefaultendpunct}{\mcitedefaultseppunct}\relax
\EndOfBibitem
\bibitem[Haehl et~al.(2012)Haehl, Evers, Grandthyll, Paulus, Sternemann,
  Loskill, Lessel, Hüsecken, Brenner, Tolan, and Jacobs]{haehl2012}
Haehl,~H.; Evers,~F.; Grandthyll,~S.; Paulus,~M.; Sternemann,~C.; Loskill,~P.;
  Lessel,~M.; Hüsecken,~A.; Brenner,~T.; Tolan,~M.; Jacobs,~K. Subsurface
  Influence on the Structure of Protein Adsorbates as Revealed by in Situ
  {X}-ray Reflectivity. \emph{Langmuir} \textbf{2012}, \emph{28},
  7747--7756\relax
\mciteBstWouldAddEndPuncttrue
\mciteSetBstMidEndSepPunct{\mcitedefaultmidpunct}
{\mcitedefaultendpunct}{\mcitedefaultseppunct}\relax
\EndOfBibitem
\bibitem[Langdon et~al.(2012)Langdon, Kastantin, and Schwartz]{Langdon2012}
Langdon,~B.~B.; Kastantin,~M.; Schwartz,~D.~K. Apparent Activation Energies
  Associated with Protein Dynamics on Hydrophobic and Hydrophilic Surfaces.
  \emph{Biophysical Journal} \textbf{2012}, \emph{102}, 2625--2633\relax
\mciteBstWouldAddEndPuncttrue
\mciteSetBstMidEndSepPunct{\mcitedefaultmidpunct}
{\mcitedefaultendpunct}{\mcitedefaultseppunct}\relax
\EndOfBibitem
\bibitem[Ho et~al.(2008)Ho, Danquah, Wang, and Forde]{Ho_2008}
Ho,~J.; Danquah,~M.~K.; Wang,~H.; Forde,~G.~M. Protein loaded mesoporous silica
  spheres as a controlled delivery platform. \emph{J. Chem. Technol.
  Biotechnol.} \textbf{2008}, \emph{83}, 2426--2433\relax
\mciteBstWouldAddEndPuncttrue
\mciteSetBstMidEndSepPunct{\mcitedefaultmidpunct}
{\mcitedefaultendpunct}{\mcitedefaultseppunct}\relax
\EndOfBibitem
\bibitem[Slowing et~al.(2007)Slowing, Trewyn, and Lin]{slowing_2007}
Slowing,~I.; Trewyn,~B.; Lin,~V. Mesoporous Silica Nanoparticles for
  Intracellular Delivery of Membrane-Impermeable Proteins. \emph{J. Am. Chem.
  Soc.} \textbf{2007}, \emph{129}, 8845--8849\relax
\mciteBstWouldAddEndPuncttrue
\mciteSetBstMidEndSepPunct{\mcitedefaultmidpunct}
{\mcitedefaultendpunct}{\mcitedefaultseppunct}\relax
\EndOfBibitem
\bibitem[Deere et~al.(2002)Deere, Magner, Wall, and Hodnett]{Deere_2002}
Deere,~J.; Magner,~E.; Wall,~J.~G.; Hodnett,~K. Mechanistic and structural
  features of protein adsorption onto mesoporous silicates. \emph{J. Phys.
  Chem. B} \textbf{2002}, \emph{106}, 532--536\relax
\mciteBstWouldAddEndPuncttrue
\mciteSetBstMidEndSepPunct{\mcitedefaultmidpunct}
{\mcitedefaultendpunct}{\mcitedefaultseppunct}\relax
\EndOfBibitem
\bibitem[Hudson et~al.(2005)Hudson, Magner, Cooney, and Hodnett]{Hudson_2005}
Hudson,~S.; Magner,~E.; Cooney,~J.; Hodnett,~B.~K. Methodology for the
  immobilization of enzymes onto mesoporous materials. \emph{J. Phys. Chem. B}
  \textbf{2005}, \emph{109}, 19496--19506\relax
\mciteBstWouldAddEndPuncttrue
\mciteSetBstMidEndSepPunct{\mcitedefaultmidpunct}
{\mcitedefaultendpunct}{\mcitedefaultseppunct}\relax
\EndOfBibitem
\bibitem[Miyahara et~al.(2006)Miyahara, Vinu, Hossain, Nakanishi, and
  Ariga]{miyahara_2006_2}
Miyahara,~M.; Vinu,~A.; Hossain,~K.~Z.; Nakanishi,~T.; Ariga,~K. Adsorption
  study of heme proteins on {SBA-15} mesoporous silica with pore-filling
  models. \emph{Thin Solid Films} \textbf{2006}, \emph{499}, 13--18\relax
\mciteBstWouldAddEndPuncttrue
\mciteSetBstMidEndSepPunct{\mcitedefaultmidpunct}
{\mcitedefaultendpunct}{\mcitedefaultseppunct}\relax
\EndOfBibitem
\bibitem[Vinu et~al.(2004)Vinu, Murugesan, Tangermann, and Hartmann]{Vinu_2004}
Vinu,~A.; Murugesan,~V.; Tangermann,~O.; Hartmann,~M. Adsorption of cytochrome
  \textit{c} on mesoporous molecular sieves: Influence of p{H}, pore diameter,
  and aluminum incorporation. \emph{Chem. Mater.} \textbf{2004}, \emph{16},
  3056--3065\relax
\mciteBstWouldAddEndPuncttrue
\mciteSetBstMidEndSepPunct{\mcitedefaultmidpunct}
{\mcitedefaultendpunct}{\mcitedefaultseppunct}\relax
\EndOfBibitem
\bibitem[Zhang et~al.(2007)Zhang, Wang, Wu, Qian, and Man]{zhang_2007}
Zhang,~X.; Wang,~J.; Wu,~W.; Qian,~S.; Man,~Y. Immobilization and
  electrochemistry of cytochrome \textit{c} on amino-functionalized mesoporous
  silica thin films. \emph{Electrochem. Commun.} \textbf{2007}, \emph{9},
  2098--2104\relax
\mciteBstWouldAddEndPuncttrue
\mciteSetBstMidEndSepPunct{\mcitedefaultmidpunct}
{\mcitedefaultendpunct}{\mcitedefaultseppunct}\relax
\EndOfBibitem
\bibitem[Hartmann(2005)]{Hartmann_2005}
Hartmann,~M. Ordered mesoporous materials for bioadsorption and biocatalysis.
  \emph{Chem. Mater.} \textbf{2005}, \emph{17}, 4577--4593\relax
\mciteBstWouldAddEndPuncttrue
\mciteSetBstMidEndSepPunct{\mcitedefaultmidpunct}
{\mcitedefaultendpunct}{\mcitedefaultseppunct}\relax
\EndOfBibitem
\bibitem[Wu and Zhao(2011)Wu, and Zhao]{zhao_2011}
Wu,~Z.; Zhao,~D. Ordered mesoporous materials as adsorbents. \emph{Chem.
  Commun.} \textbf{2011}, \emph{47}, 3332--3338\relax
\mciteBstWouldAddEndPuncttrue
\mciteSetBstMidEndSepPunct{\mcitedefaultmidpunct}
{\mcitedefaultendpunct}{\mcitedefaultseppunct}\relax
\EndOfBibitem
\bibitem[Essa et~al.(2007)Essa, Magner, Cooney, and Hodnett]{Essa_2007}
Essa,~H.; Magner,~E.; Cooney,~J.; Hodnett,~B.~K. Influence of p{H} and ionic
  strength on the adsorption, leaching and activity of myoglobin immobilized
  onto ordered mesoporous silicates. \emph{J. Mol. Catal. B: Enzym.}
  \textbf{2007}, \emph{49}, 61--68\relax
\mciteBstWouldAddEndPuncttrue
\mciteSetBstMidEndSepPunct{\mcitedefaultmidpunct}
{\mcitedefaultendpunct}{\mcitedefaultseppunct}\relax
\EndOfBibitem
\bibitem[Hartvig et~al.(2011)Hartvig, van~de Weert, Ostergaard, and
  Jensen]{Hartvig_2011}
Hartvig,~R.~A.; van~de Weert,~M.; Ostergaard,~J.; Jensen,~H. Protein Adsorption
  at Charged Surfaces: The Role of Electrostatic Interactions and Interfacial
  Charge Regulation. \emph{Langmuir} \textbf{2011}, \emph{27}, 2634--2643\relax
\mciteBstWouldAddEndPuncttrue
\mciteSetBstMidEndSepPunct{\mcitedefaultmidpunct}
{\mcitedefaultendpunct}{\mcitedefaultseppunct}\relax
\EndOfBibitem
\bibitem[Pasche et~al.(2005)Pasche, Vörös, Griesser, Spencer, and
  Textor]{pasche_2005}
Pasche,~S.; Vörös,~J.; Griesser,~H.~J.; Spencer,~N.~D.; Textor,~M. Effects of
  Ionic Strength and Surface Charge on Protein Adsorption at {PEG}ylated
  Surfaces. \emph{J. Phys. Chem. B} \textbf{2005}, \emph{109},
  17545--17552\relax
\mciteBstWouldAddEndPuncttrue
\mciteSetBstMidEndSepPunct{\mcitedefaultmidpunct}
{\mcitedefaultendpunct}{\mcitedefaultseppunct}\relax
\EndOfBibitem
\bibitem[Qiao et~al.(2005)Qiao, Yu, Xing, Hu, Djojoputro, and Lu]{qiao_2005}
Qiao,~S.~Z.; Yu,~C.; Xing,~W.; Hu,~Q.~H.; Djojoputro,~H.; Lu,~G.~Q. Synthesis
  and bio-adsorptive properties of large-pore periodic mesoporous organosilica
  rods. \emph{Chem. Mater.} \textbf{2005}, \emph{17}, 6172--6176\relax
\mciteBstWouldAddEndPuncttrue
\mciteSetBstMidEndSepPunct{\mcitedefaultmidpunct}
{\mcitedefaultendpunct}{\mcitedefaultseppunct}\relax
\EndOfBibitem
\bibitem[Meier-Koll et~al.(2004)Meier-Koll, Fleck, and von
  Grünberg]{meier-koll_2004}
Meier-Koll,~A.; Fleck,~C.; von Grünberg,~H. The counterion-release interaction.
  \emph{J. Phys.: Condens. Matter} \textbf{2004}, \emph{16}, 6041--6052\relax
\mciteBstWouldAddEndPuncttrue
\mciteSetBstMidEndSepPunct{\mcitedefaultmidpunct}
{\mcitedefaultendpunct}{\mcitedefaultseppunct}\relax
\EndOfBibitem
\bibitem[Causserand et~al.(2001)Causserand, Kara, and Aimar]{causserand_2001}
Causserand,~C.; Kara,~Y.; Aimar,~P. Protein fractionation using selective
  adsorption on clay surface before filtration. \emph{J. Membr. Sci.}
  \textbf{2001}, \emph{186}, 165--181\relax
\mciteBstWouldAddEndPuncttrue
\mciteSetBstMidEndSepPunct{\mcitedefaultmidpunct}
{\mcitedefaultendpunct}{\mcitedefaultseppunct}\relax
\EndOfBibitem
\bibitem[Kraning et~al.(2007)Kraning, Benz, Bloome, Campanello, Fahrenbach,
  Mistry, Hedge, Clevenger, Gligorich, Hopkins, Hoops, Mendes, Chang, and
  Su]{kraning_2007}
Kraning,~C.~M.; Benz,~T.~L.; Bloome,~K.~S.; Campanello,~G.~C.;
  Fahrenbach,~V.~S.; Mistry,~S.~A.; Hedge,~C.~A.; Clevenger,~K.~D.;
  Gligorich,~K.~M.; Hopkins,~T.~A.; Hoops,~G.~C.; Mendes,~S.~B.; Chang,~H.-C.;
  Su,~M.-C. Determination of surface coverage and orientation of reduced
  cytochrome \textit{c} on a silica surface with polarized ATR Spectroscopy.
  \emph{J. Phys. Chem. C} \textbf{2007}, \emph{111}, 13062--13067\relax
\mciteBstWouldAddEndPuncttrue
\mciteSetBstMidEndSepPunct{\mcitedefaultmidpunct}
{\mcitedefaultendpunct}{\mcitedefaultseppunct}\relax
\EndOfBibitem
\bibitem[Welsch et~al.(2012)Welsch, Becker, Dzubiella, and
  Ballauff]{welsch_2012}
Welsch,~N.; Becker,~A.~L.; Dzubiella,~J.; Ballauff,~M. Core-shell microgels as
  smart carriers for enzymes. \emph{Soft Matter} \textbf{2012}, \emph{8},
  1428--1436\relax
\mciteBstWouldAddEndPuncttrue
\mciteSetBstMidEndSepPunct{\mcitedefaultmidpunct}
{\mcitedefaultendpunct}{\mcitedefaultseppunct}\relax
\EndOfBibitem
\bibitem[Yigit et~al.(2012)Yigit, Welsch, Ballauff, and
  Dzubiella]{yigit_welsch}
Yigit,~C.; Welsch,~N.; Ballauff,~M.; Dzubiella,~J. Protein Sorption to Charged
  Microgels: Characterizing Binding Isotherms and Driving Forces.
  \emph{Langmuir} \textbf{2012}, \emph{28}, 14373--14385\relax
\mciteBstWouldAddEndPuncttrue
\mciteSetBstMidEndSepPunct{\mcitedefaultmidpunct}
{\mcitedefaultendpunct}{\mcitedefaultseppunct}\relax
\EndOfBibitem
\bibitem[Tsong(1975)]{tsong_1975}
Tsong,~T.~Y. An Acid Induced Conformational Transition of Denatured Cytochrome
  \textit{c} in Urea and Guanidine Hydrochloride. \emph{Biochemistry}
  \textbf{1975}, \emph{14}, 1542--1547\relax
\mciteBstWouldAddEndPuncttrue
\mciteSetBstMidEndSepPunct{\mcitedefaultmidpunct}
{\mcitedefaultendpunct}{\mcitedefaultseppunct}\relax
\EndOfBibitem
\bibitem[Hsu et~al.(2007)Hsu, Shiu, Jeng, Chen, Huang, Lai, Tsai, Jang, Lee,
  Lin, Lin, and Wang]{Hsu_2007}
Hsu,~I.-J.; Shiu,~Y.-J.; Jeng,~U.-S.; Chen,~T.-H.; Huang,~Y.-S.; Lai,~Y.-H.;
  Tsai,~L.-N.; Jang,~L.; Lee,~J.-F.; Lin,~L.-J.; Lin,~S.-H.; Wang,~Y. A
  solution study on the local and global structure changes of cytochrome
  \textit{c}: An unfolding process induced by urea. \emph{J. Phys. Chem. A}
  \textbf{2007}, \emph{111}, 9286--9290\relax
\mciteBstWouldAddEndPuncttrue
\mciteSetBstMidEndSepPunct{\mcitedefaultmidpunct}
{\mcitedefaultendpunct}{\mcitedefaultseppunct}\relax
\EndOfBibitem
\bibitem[Zhao et~al.(1998)Zhao, Feng, Huo, Melosh, Fredrickson, Chmelka, and
  Stucky]{zhao1998}
Zhao,~D.~Y.; Feng,~J.~L.; Huo,~Q.~S.; Melosh,~N.; Fredrickson,~G.~H.;
  Chmelka,~B.~F.; Stucky,~G.~D. Triblock copolymer syntheses of mesoporous
  silica with periodic 50 to 300 angstrom pores. \emph{Science} \textbf{1998},
  \emph{279}, 548--552\relax
\mciteBstWouldAddEndPuncttrue
\mciteSetBstMidEndSepPunct{\mcitedefaultmidpunct}
{\mcitedefaultendpunct}{\mcitedefaultseppunct}\relax
\EndOfBibitem
\bibitem[Hofmann et~al.(2005)Hofmann, Wallacher, Huber, Birringer, Knorr,
  Schreiber, and Findenegg]{Hofmann2005}
Hofmann,~T.; Wallacher,~D.; Huber,~P.; Birringer,~R.; Knorr,~K.; Schreiber,~A.;
  Findenegg,~G.~H. Small-angle x-ray diffraction of Kr in mesoporous silica:
  Effects of microporosity and surface roughness. \emph{Phys. Rev. B: Condens.
  Matter Mater. Phys.} \textbf{2005}, \emph{72}, 064122\relax
\mciteBstWouldAddEndPuncttrue
\mciteSetBstMidEndSepPunct{\mcitedefaultmidpunct}
{\mcitedefaultendpunct}{\mcitedefaultseppunct}\relax
\EndOfBibitem
\bibitem[Zickler et~al.(2006)Zickler, Jaehnert, Wagermaier, Funari, Findenegg,
  and Paris]{Zickler2006}
Zickler,~G.~A.; Jaehnert,~S.; Wagermaier,~W.; Funari,~S.~S.; Findenegg,~G.~H.;
  Paris,~O. Physisorbed films in periodic mesoporous silica studied by in situ
  synchrotron small-angle diffraction. \emph{Physical Review B} \textbf{2006},
  \emph{73}, 184109\relax
\mciteBstWouldAddEndPuncttrue
\mciteSetBstMidEndSepPunct{\mcitedefaultmidpunct}
{\mcitedefaultendpunct}{\mcitedefaultseppunct}\relax
\EndOfBibitem
\bibitem[Oellerich et~al.(2002)Oellerich, Wackerbarth, and
  Hildebrandt]{oellerich_2002}
Oellerich,~S.; Wackerbarth,~H.; Hildebrandt,~P. Spectroscopic characterization
  of nonnative conformational states of cytochrome \textit{c}. \emph{J. Phys.
  Chem. B} \textbf{2002}, \emph{106}, 6566--6580\relax
\mciteBstWouldAddEndPuncttrue
\mciteSetBstMidEndSepPunct{\mcitedefaultmidpunct}
{\mcitedefaultendpunct}{\mcitedefaultseppunct}\relax
\EndOfBibitem
\bibitem[McLendon and Smith(1978)McLendon, and Smith]{mclendon_1978}
McLendon,~G.; Smith,~M. Equilibrium and Kinetic Studies of Unfolding of
  Homologous Cytochromes c. \emph{J. Biol. Chem.} \textbf{1978}, \emph{253},
  4004--4008\relax
\mciteBstWouldAddEndPuncttrue
\mciteSetBstMidEndSepPunct{\mcitedefaultmidpunct}
{\mcitedefaultendpunct}{\mcitedefaultseppunct}\relax
\EndOfBibitem
\bibitem[Moerz et~al.(2012)Moerz, Knorr, and Huber]{moerz_2012}
Moerz,~S.~T.; Knorr,~K.; Huber,~P. Capillary condensation, freezing, and
  melting in silica nanopores: {A} sorption isotherm and scanning calorimetry
  study on nitrogen in mesoporous {SBA}-15. \emph{Phys. Rev. B: Condens. Matter
  Mater. Phys.} \textbf{2012}, \emph{85}\relax
\mciteBstWouldAddEndPuncttrue
\mciteSetBstMidEndSepPunct{\mcitedefaultmidpunct}
{\mcitedefaultendpunct}{\mcitedefaultseppunct}\relax
\EndOfBibitem
\bibitem[Cole and Saam(1974)Cole, and Saam]{sc1974}
Cole,~M.~W.; Saam,~W.~F. Excitation Spectrum and Thermodynamic Properties of
  Liquid-Film in Cylindrical Pores. \emph{Phys. Rev. Lett.} \textbf{1974},
  \emph{32}, 985--988\relax
\mciteBstWouldAddEndPuncttrue
\mciteSetBstMidEndSepPunct{\mcitedefaultmidpunct}
{\mcitedefaultendpunct}{\mcitedefaultseppunct}\relax
\EndOfBibitem
\bibitem[Saam and Cole(1975)Saam, and Cole]{sc1975}
Saam,~W.~F.; Cole,~M.~W. Excitations and Thermodynamics for Liquid-Helium
  Films. \emph{Phys. Rev. B: Solid State} \textbf{1975}, \emph{11},
  1086--1105\relax
\mciteBstWouldAddEndPuncttrue
\mciteSetBstMidEndSepPunct{\mcitedefaultmidpunct}
{\mcitedefaultendpunct}{\mcitedefaultseppunct}\relax
\EndOfBibitem
\bibitem[Imperor-Clerc et~al.(2000)Imperor-Clerc, Davidson, and
  Davidson]{imperorclerc2000}
Imperor-Clerc,~M.; Davidson,~P.; Davidson,~A. Existence of a Microporous Corona
  around the Mesopores of Silica-Based {SBA-15} Materials Templated by Triblock
  Copolymers. \emph{J. Am. Chem. Soc.} \textbf{2000}, \emph{122},
  11925--11933\relax
\mciteBstWouldAddEndPuncttrue
\mciteSetBstMidEndSepPunct{\mcitedefaultmidpunct}
{\mcitedefaultendpunct}{\mcitedefaultseppunct}\relax
\EndOfBibitem
\bibitem[Cheng et~al.(2003)Cheng, Lin, and Chang]{cheng_2003}
Cheng,~Y.-Y.; Lin,~S.~H.; Chang,~H.-C. Probing Adsorption, Orientation and
  Conformational Changes of Cytochrome \textit{c} on Fused Silica Surfaces with
  the Soret Band. \emph{J. Phys. Chem. A} \textbf{2003}, \emph{107},
  10687--10694\relax
\mciteBstWouldAddEndPuncttrue
\mciteSetBstMidEndSepPunct{\mcitedefaultmidpunct}
{\mcitedefaultendpunct}{\mcitedefaultseppunct}\relax
\EndOfBibitem
\bibitem[Miyahara et~al.(2006)Miyahara, Vinu, and Ariga]{miyahara_2006}
Miyahara,~M.; Vinu,~A.; Ariga,~K. Adsorption myoglobin over mesoporous silica
  molecular sieves: Pore Size effect and pore-filling model. \emph{Mater. Sci.
  Eng., C} \textbf{2006}, \emph{27}, 232--236\relax
\mciteBstWouldAddEndPuncttrue
\mciteSetBstMidEndSepPunct{\mcitedefaultmidpunct}
{\mcitedefaultendpunct}{\mcitedefaultseppunct}\relax
\EndOfBibitem
\bibitem[Menaa et~al.(2008)Menaa, Torres, Herrero, Rives, r.~W.~Gilbert, and
  Eggers]{menaa_2008}
Menaa,~B.; Torres,~C.; Herrero,~M.; Rives,~V.; r.~W.~Gilbert,~A.; Eggers,~D.~K.
  Protein adsorption onto organically modified silica glass leads to a
  different structure than sol-gel encapsulation. \emph{Biophys. J.}
  \textbf{2008}, \emph{95}, L51--L53\relax
\mciteBstWouldAddEndPuncttrue
\mciteSetBstMidEndSepPunct{\mcitedefaultmidpunct}
{\mcitedefaultendpunct}{\mcitedefaultseppunct}\relax
\EndOfBibitem
\bibitem[Tanford et~al.(1966)Tanford, Kawahara, and Lapanje]{kawahara_1966_2}
Tanford,~C.; Kawahara,~K.; Lapanje,~S. Proteins in 6 {M} Guanidine
  Hydrochloride - Demonstration of Random Coil Behavior. \emph{J. Biol. Chem.}
  \textbf{1966}, \emph{241}, 1921--1923\relax
\mciteBstWouldAddEndPuncttrue
\mciteSetBstMidEndSepPunct{\mcitedefaultmidpunct}
{\mcitedefaultendpunct}{\mcitedefaultseppunct}\relax
\EndOfBibitem
\bibitem[Becker et~al.(2012)Becker, Henzler, Welsch, Ballauff, and
  Borisov]{becker_welsch_2012}
Becker,~A.~L.; Henzler,~K.; Welsch,~N.; Ballauff,~M.; Borisov,~O. Proteins and
  polyelectrolytes: A charged relationship. \emph{Curr. Opin. Colloid Interface
  Sci.} \textbf{2012}, \emph{17}, 90--96\relax
\mciteBstWouldAddEndPuncttrue
\mciteSetBstMidEndSepPunct{\mcitedefaultmidpunct}
{\mcitedefaultendpunct}{\mcitedefaultseppunct}\relax
\EndOfBibitem
\bibitem[Perroud et~al.(2005)Perroud, Bokoch, and Zare]{Perroud2005}
Perroud,~T.~D.; Bokoch,~M.~P.; Zare,~R.~N. Cytochrome c conformations resolved
  by the photon counting histogram: Watching the alkaline transition with
  single-molecule sensitivity. \emph{Proceedings of the National Academy of
  Sciences of the United States of America} \textbf{2005}, \emph{102},
  17570--17575\relax
\mciteBstWouldAddEndPuncttrue
\mciteSetBstMidEndSepPunct{\mcitedefaultmidpunct}
{\mcitedefaultendpunct}{\mcitedefaultseppunct}\relax
\EndOfBibitem
\end{mcitethebibliography}
\bibliographystyle{achemso.bst}
\providecommand*\mcitethebibliography{\thebibliography}
\csname @ifundefined\endcsname{endmcitethebibliography}
  {\let\endmcitethebibliography\endthebibliography}{}

\end{document}